\documentstyle[twocolumn,aps,prl]{revtex}
\begin{document}
\ifpreprintsty\else
\twocolumn[\hsize\textwidth%
\columnwidth\hsize\csname@twocolumnfalse\endcsname
\fi
\draft
\preprint{ }
\title {Time-dependent density functional theory beyond the adiabatic
local density approximation} \author {G. Vignale, C. A. Ullrich} \address
{Department of Physics,  University of Missouri, Columbia,  Missouri
65211} \author {S. Conti} \address {Scuola Normale Superiore,
I-56126 Pisa, Italy}
\date{\today} 
\maketitle

\begin{abstract}
In the current density functional theory of 
linear and nonlinear time-dependent  phenomena, the treatment of exchange and
correlation beyond the level of the 
adiabatic local density approximation  is shown to lead to the
appearance  of viscoelastic stresses in the electron fluid.  Complex and
frequency-dependent viscosity/elasticity coefficients are microscopically
derived and expressed in terms of properties of the homogeneous electron gas.
As a first consequence of this formalism, we provide an explicit formula for
the linewidths of collective excitations in electronic systems.
\end{abstract}

\pacs{71.45Gm;73.20Dx,Mf;78.30Fs;21.10Re;36.40+d;85.42+m}
\ifpreprintsty\else\vskip1pc]\fi
\narrowtext

Time-dependent density functional theory (TDFT) \cite {Gross}
is frequently invoked as a tool for studying the
dynamics of many-particle systems.  This theory maps the
difficult problem of interacting electrons in a time-dependent external
potential $V(\vec r,t)$  to the  simpler one of noninteracting
electrons in an effective time-dependent potential $V_{eff}(\vec r, t) =
V(\vec r,t) + v_H(\vec r,t) + v_{xc}(\vec r,t)$ (where $v_H$ is the
Hartree potential, and $v_{xc}$  is the exchange-correlation (xc)
potential) yielding the same density $n(\vec r, t)$.  In order to obtain
a practical computational scheme, the xc potential is usually
 approximated as a function of the  instantaneous local density,
\begin {equation}
v_{xc}^{ALDA}(n(\vec r, t)) = \left ({d \epsilon_{xc} (n) \over d
n}\right )_{n=  n(\vec r,t)}, \label {ALDA} \end {equation} where
$\epsilon_{xc} (n)$ is the xc energy density of the homogeneous electron
gas of density $n$.  This scheme is known as the ``adiabatic local
density approximation" (ALDA)\cite {Zangwill}.  Because the conditions
of validity of the ALDA (slowly varying density and potential in {\it
time} as well as in space) are seldom met in experiments,  a few
attempts  have been made \cite{GK,Dobson} to improve upon the ALDA. 
The objective of these attempts was to obtain  approximations for
the xc potential which would still be local in space, but  not in time.
All these approximations  were found to suffer from inconsistencies,
such as the failure to satisfy the so called ``harmonic potential
theorem" (HPT) \cite{Dobson,Vignale,Kohn},  or other basic symmetries. 
Only recently, it has become clear that the root of these difficulties
lies in the fact that the xc potential in TDFT is an intrinsically nonlocal
functional of the density, that is, a functional that does not admit a
gradient expansion \cite{VK1,VK2}.
 
Fortunately, a local gradient expansion is still possible if the theory
is formulated in terms of the {\it current density}. This was shown 
in ref. \cite{VK1} by Vignale and Kohn, who developed the time-dependent
current density functional approach to the linear response theory,  and
gave an explicit expression for the linearized  xc vector potential $\vec
a_{xc} (\vec r, \omega)$   for a system of slowly varying density, 
subject to a spatially slowly varying external vector potential  at a
finite frequency $\omega$. Their expression becomes exact in the
limit $k\ll  
\omega/v_F,k_F$ and $q\ll  \omega/v_F,k_F$, where $k^{-1}$ and $q^{-1}$
are the characteristic length scales for variation of the external
potential and equilibrium density, respectively, and $k_F$ and $v_F$ are
the local Fermi momentum and velocity. However, the final expression for $\vec
a_{xc} (\vec r,\omega)$ in ref.\cite{VK1}  is rather formidable,  and its
physical meaning is far from transparent.  Furthermore,  it is
restricted to the linear response regime. It is the purpose of this paper to
overcome these limitations.  

In the following we derive  a  consistent local
theory of the nonlinear dynamical response of a quantum electronic 
system of ``slowly varying density", in the sense specified above.  The
effect of the xc potential beyond the ALDA will be
shown to be  analogous to the introduction of  viscoelastic stresses in
classical fluid dynamics and elasticity theory \cite {Landau}.  The 
generalized  viscosity coefficients (or, equivalently, the generalized
bulk and shear moduli) are complex and frequency-dependent functions of
the density, and can be calculated in terms of the local field factors
of the uniform electron gas \cite{Conti,Nifosi}.
An important  consequence of the xc viscosity is to
provide a damping mechanism for long-lived collective excitations
which cannot effectively decay into particle-hole pairs -  the only form of
damping allowed within the ALDA \cite{Zaremba}.

We begin by recasting the linear response theory of \cite{VK1,VK2} 
in a form that is suitable for the nonlinear generalization.
Let $n_0(\vec r)$ be the ground-state density of the system, and let 
${\chi}_{KS,ij} (\vec r,\vec r',\omega)$ be the current-current
response function of a system of noninteracting electrons whose
ground-state density is also $n_0(\vec r)$. The system is perturbed by
a weak external vector potential $\vec a_1(\vec r, t) = 
\vec a_1(\vec r, \omega) e^{-i \omega t}$ \cite{Footnote1}, and we want
to calculate the amplitude of the current-density response 
 $\vec j_1(\vec r, t) =  \vec j_1(\vec r, \omega) e^{-i \omega t}$, to
first order in $\vec a_1$.  The answer is
\begin {eqnarray}
j_{1,i}(\vec r,\omega) &=& \int \sum_j {\chi}_{KS,ij} (\vec r,\vec
r',\omega) \left[  a_{1,j}(\vec r',\omega) \right.\nonumber\\
&& \left.+   a_{H1,j}(\vec r',\omega)+ a_{xc1,j}(\vec r',\omega) \right]
d\vec r', \label {KohnSham}
\end {eqnarray}
where $\vec a_{H1}$ is the first-order change in the Hartree potential
(written in vector-potential form \cite{Footnote1}), and $\vec a_{xc1}$
is the first-order xc vector potential, which contains the many-body
effects. Note that $\vec a_{xc1}$ in general has both longitudinal and
transverse components, even  when $\vec a_1$ is purely longitudinal. 
The local-density approximation for $\vec a_{xc1}$ was first derived in 
ref. {\cite {VK1}}, eq.~(19).  We have found that that complicated
formula can be written in a physically transparent form.  To this end, 
we introduce the xc ``electric field" $\vec E_{xc1}
(\vec r, \omega) \equiv  {i \omega \over c} \vec a_{xc1}(\vec r,
\omega)$. Then  
\begin {eqnarray}  -e
E_{xc1,i} (\vec r, \omega) &=& - \vec \nabla_i v_{xc1}^{ALDA}(\vec r
,\omega)  \nonumber\\
&& +{1 \over n_0(\vec r)} \sum_j {\partial \sigma_{xc,ij} (\vec r,
\omega) \over \partial r_j}
\label {exc}  \end {eqnarray}
($e$ is the absolute value of the electron charge, and $c$ is the speed
of light).  The first term is  the linearization of the ALDA
expression (\ref {ALDA}), and the dynamical correction is the
divergence of the visco-elastic stress tensor  \begin {equation}
\sigma_{xc,ij} = \tilde \eta_{xc}\left({\partial u_i \over \partial r_j}
+ {\partial u_j \over \partial r_i} - {2 \over 3} \vec \nabla \cdot \vec
u \delta_{ij}\right) + \tilde \zeta_{xc} \vec \nabla \cdot \vec u
\delta_{ij}. \label {stress} \end {equation}
Here $\vec u (\vec r, \omega) \equiv \vec
j_1 (\vec r,\omega)/n_0(\vec r)$ is the velocity field, and  $\tilde
\eta_{xc} (\omega,n_0(\vec r))$ and $\tilde \zeta_{xc}
(\omega, n_0(\vec r))$  are complex  viscosity coefficients.  They
are related to the homogeneous electron gas functions
$f_{xc,L}^h(\omega, n)$ and $f_{xc,T}^h(\omega, n)$
($L$ for longitudinal, $T$ for transverse) used in ref. \cite {VK1} as
follows:
\begin {equation}
\tilde \zeta_{xc}(\omega,n) = - {n^2 \over i \omega}
\left[f_{xc,L}^h- \!{4 \over 3}f_{xc,T}^h
-\! {d^2 \epsilon_{xc}(n) \over dn^2}\right]\label
{zetaxc} \end {equation} 
and 
\begin {equation}
\tilde \eta_{xc}(\omega,n) = - {n^2 \over i \omega}
f_{xc,T}^h\,.
 \label {etaxc} \end {equation}
The functions $f_{xc,L(T)}^h(\omega, n)$  
are  defined in terms  of the dynamical local
field factors $G_{L(T)}(k,\omega)$\cite{Tosi} as
$f_{xc,L(T)}^h(\omega, n) \equiv -\lim_{k 
\to 0} 4 \pi e^2 G_{L(T)}(k,\omega)/k^2$.
They have been recently
calculated, within a mode-coupling approximation scheme, at zero temperature,
by Conti {\em et al.} \cite{Conti,Nifosi}.  
Because the local field factors are singular for small
$k$ and $\omega$, it is important to keep track of the  order of the
limits $k \to 0$ and $\omega \to
 0$.  The $f^h_{xc}$'s are defined by taking the limit $k \to 0$ first.
Thus the limit for $\omega \to 0$ of  $f_{xc,L}^h(\omega,
n)$  differs from the familiar   $\lim_{k \to 0} 4 \pi e^2 G_{L}(k,0)/k^2
= -d^2 \epsilon_{xc}(n) / dn^2$.  However, {\it rigorous}
 low frequency limits \cite {Footnote0} can  be obtained
\cite{Conti2} from an analysis of the transport equation in the  Landau theory
of Fermi liquids \cite{Nozieres}.  We find
\begin {equation}
\lim_{\omega \to 0} {\rm Re}\left[f_{xc,L}^h(\omega, n)-  {4 \over
3}f_{xc,T}^h(\omega,n) - {d^2 \epsilon_{xc}(n) \over dn^2}\right] = 0
\label {lim1} \end {equation}
and  
\begin {equation}
\lim_{\omega \to 0} {\rm Re}f_{xc,T}^h(\omega, n) = {2E_F \over 5 n}
{F_2/5-F_1/3 \over 1 + F_1/3}, \label {lim2} \end {equation}
where $F_0, F_1...$ are the usual dimensionless Landau parameters of
the homogeneous electron gas \cite{Nozieres}. As for the
imaginary parts, one finds ${\rm Im} f_{xc,L(T)}^h \sim -c_{0,L(T)} \omega$ 
for  $\omega \to 0$, where the approximate values of the coefficients
$c_{0,L(T)}$ are tabulated in ref.~\cite{Nifosi}.

 From these results, one  concludes that the real parts of the xc
viscosity coefficients (which agree with the ordinary notion of fluid
viscosities) have finite values in the limit $\omega \to 0$. The
imaginary parts of the viscosity coefficients are better understood in
terms of bulk and shear moduli of an isotropic elastic medium,
$K^{dyn}_{xc}$ and $\mu^{dyn}_{xc}$.
According to elasticity theory \cite
{Landau}, we  define $K^{dyn}_{xc} (\omega) = \omega {\rm Im} \tilde
\zeta_{xc}$ and $\mu^{dyn}_{xc} (\omega) = \omega {\rm Im} \tilde
\eta_{xc}$. The superscript {\it dyn} is  
a reminder that these are  {\it dynamical} contributions to  be added to the
usual static ones,  already present in the ALDA. The static elastic constants
are  $K^{stat}_{xc} = n^2 d^2 \epsilon_{xc}(n)/dn^2$ and $\mu^{stat}_{xc} = 0$
respectively.  
Equations (\ref {lim1},\ref {lim2}) show that for $\omega \to 0$
$K^{dyn}_{xc}$ vanishes,  while $\mu^{dyn}_{xc}$ has a finite value. 
A similar state of affairs holds for the  noninteracting {\it kinetic}
contributions to the bulk and shear moduli: 
$K^{dyn}_{kin} = 0$ and   
$\mu^{dyn}_{kin} = p(n)$, where $p(n)$ is the noninteracting Fermi
pressure (see below), and $K^{stat}_{kin} = n dp(n)/dn$,
$\mu^{stat}_{kin} = 0$.   The general conclusion is that dynamical
(post-ALDA) effects do not modify the bulk modulus,  but they cause the
appearance of a nonvanishing shear modulus and viscosity.

Equations (\ref{exc}) and (\ref{stress}) clearly  display the
basic symmetries which were used in the derivation of \cite{VK1,VK2}.
First of all, the fact that the force exerted by the xc potential per
unit volume 
$-e n_0(\vec r) \vec E_{xc1}(\vec r, \omega) -
n_1(\vec r,\omega) \vec \nabla v_{xc0}(\vec r)$ can be written as the
divergence of a  symmetric  rank 2 local tensor guarantees that the net
force and the net torque acting on a volume element of the fluid have no
contribution from the volume element itself (Newton's third
law). Therefore, the HPT, the ``zero force''  and ``zero-torque''
theorems of \cite{VK1,VK2} are manifestly satisfied. (Note that the
force exerted by the xc ``magnetic field'' $\vec B_{xc} = \vec \nabla
\times \vec a_{xc}$ is legitimately disregarded in this argument,  being of
higher order in the gradient expansion [see also below]. Besides, it is
rigorously absent in the linear response theory, if there is no static
magnetic field).

Thus far, we have used the condition of slow density variation ($k, q
\ll  \omega/v_F$) only in approximating the xc vector potential.
If this condition is met in the physical system under study, we can {\it
also} use it to approximate the Kohn-Sham (KS) response function,
${\chi}_{KS,ij} (\vec r, \vec r',\omega)$. Then eq.~(\ref{KohnSham}) for
the current  reduces (after considerable algebraic manipulation) to a
linearized Navier-Stokes equation \cite{Landau} with complex and
frequency-dependent viscosity coefficients:  
\begin {eqnarray} -i
m\omega j_{1,i}(\vec r, \omega) &=& n_0(\vec r) \left[-i \omega
{e \over c} a_i(\vec r,\omega) \right. \nonumber\\
&&\hskip-10mm\left.
- \nabla_i \left ({p_1(n)
\over n_0} + v_{H1}(\vec r, \omega) +  v_{xc1}^{ALDA}(\vec r,
\omega)\right)\right]\nonumber \\
&& +  \sum_j {\partial \sigma_{ij}(\vec
r,\omega) \over \partial r_j}. \label {Navierstokes} \end {eqnarray}
Here $p_1(n) = p(n)-p(n_0)$ is the first order change in the
pressure of the noninteracting electron fluid  ($p(n)
= (3 \pi^2)^{2/3} \hbar^2 n^{5/3}/5m$), and  $v_{H1}$,
$ v_{xc1}^{ALDA}$ are the
first-order changes in the Hartree potential and ALDA xc potential.  The
{\it full} stress tensor $\sigma_{ij}$  is defined as in eq.~(\ref
{stress}),  with the viscosity coefficient $\tilde \eta_{xc}$ replaced by
$\tilde \eta_{xc} - p(n_0)/i \omega$, while $\tilde \zeta_{xc}$ remains
unchanged.

If, on the other hand, the conditions  $k,q\ll  \omega /v_F$ are not
well satisfied by our system, then it is better to revert to the original
KS formulation (\ref{KohnSham}), which treats the
non-interacting response exactly.  The use of the local density
approximation (\ref{exc}) for the xc potential becomes then an
uncontrolled approximation, but it may still work well in practice. In
particular, we note that eq.~(\ref{KohnSham}) allows for the phenomenon of
Landau damping  (damping of collective modes by single electron-hole pairs), 
while eq.~(\ref{Navierstokes}) does not.

 Let us now discuss the
generalization of the formalism to the {\it
nonlinear} response regime. In this case, one must solve the full
time-dependent KS equations for the KS orbitals
$\psi_\alpha (\vec r,t)$
\begin {eqnarray}
\left[
\vbox to \ifpreprintsty  0.85\baselineskip{} \else 1.2\baselineskip{} \fi
i \hbar{\partial \over \partial t}- {1 \over 2m}\left(- i \hbar
\vec \nabla + 
{e \over c}  \vec a(\vec r,t)+ {e \over c} \vec a_{xc}(\vec r,t)\right)^2
\right.\nonumber&&\\
\left.
-v_0(\vec r) -v_H(r,t)
\vbox to \ifpreprintsty  0.85\baselineskip{} \else 1.2\baselineskip{} \fi
\right]\psi_\alpha (\vec r,t) &=&0,
\label{tdks} \end {eqnarray} 
starting, for example,  with the static
KS orbitals corresponding to the external potential $v_0(\vec
r)$ at the initial time. The density and the current
density are
 computed from the KS orbitals according  to the usual rules
\cite{GK}.    The form of the nonlinear xc vector potential is dictated by the
following requirements:  (i) the xc  {\it force} density 
\begin{equation}F_{xc,i} = n{e\over c} \left[\left({\partial \over
\partial t} + \vec u \cdot 
\vec \nabla\right)a_{xc,i} -\sum_k u_k \nabla_i a_{xc,k}\right]
\label{fxc} \end 
{equation} must  be the divergence of a local symmetric rank two stress tensor 
(Newton's third law).  `Locality' here  means that $\sigma_{xc,ij}(\vec
r,t)$ is a function of $n(\vec R,t')$, $\vec j(\vec R,t')$ and their spatial 
derivatives, where $t'<t$, and $\vec 
R(t'\vert \vec r , t)$ is the position at time $t'$ of the fluid element which
evolves into $\vec r$ at time $t$ \cite{Bunner}.  (ii) Under
transformation to an accelerated frame of reference \cite{Vignale}
with origin at $\vec x(t)$, the stress tensor 
$\sigma_{xc,ij}(\vec r,t)$ becomes $\sigma'_{xc,ij}(\vec r,t) = 
\sigma_{xc,ij}(\vec r +\vec x(t),t) $. (iii) Equation
(\ref{tdks})  must reduce to the linear response theory in the limit of small
external perturbations, and to the {\it nonlinear} Navier-Stokes  equation
 in the limit  of slowly varying (in time) perturbation \cite{hydro}.

To within the accuracy of our approximation, i.e.,  {\it  to second order
in the spatial derivatives}, the above requirements 
uniquely  determine the form of  $\vec a_{xc}$:
 \begin
{equation}  {e \over c}  {\partial a_{xc,i}(\vec r,t) \over \partial
t}=
 -\! \nabla_i v_{xc}^{ALDA}(\vec r,t)  \!+\! {1 \over n(\vec r,t)}\! \sum_j
{\partial \sigma_{xc,ij} (\vec r, t) \over \partial r_j},
\label {axcnl}  \end {equation}
where
\begin {eqnarray}
\sigma_{ij} (\vec r,t) &\!=\!&  \int_{-\infty}^{t} \left[
\vbox to \ifpreprintsty  0.85\baselineskip{} \else 1.2\baselineskip{} \fi
\tilde \eta (n(\vec r
,t), t-t') \left({\partial u_i (\vec r,t') \over \partial r_j} + {\partial u_j
(\vec r,t') \over \partial r_i} \right.\right.
\nonumber\\
&&\left.\left. - {2 \over 3} \vec \nabla
\cdot \vec u (\vec r,t') \delta_{ij}\right) \right. \nonumber \\&&  \left. 
+ \tilde \zeta (n(\vec r ,t),t-t') \vec
\nabla \cdot \vec u (\vec r,t')  \delta_{ij}
\vbox to \ifpreprintsty  0.85\baselineskip{} \else 1.2\baselineskip{} \fi
\right]  dt', \label
{stressnl} \end {eqnarray}
 $\tilde \eta  (n,t-t') \equiv \int \tilde \eta 
(n,\omega) \exp (-i \omega (t-t'))d \omega /2\pi$, and similarly for $\tilde
\zeta$. Here $n(\vec r, t)$ and $\vec u(\vec r,t)$ are the time-dependent
values of the density and velocity field.

Note that our formula for   $\vec a_{xc}$ is still 
linear   in $\vec u(\vec r,t)$.  This happens
because, due to the constraint (ii) of generalized galilean
invariance,  the velocity must enter the stress tensor through its
spatial derivatives, which are assumed to be small, even if the velocity
itself is not small.  Terms of higher order in the velocity would
necessarily be of higher order in the gradient expansion. For the same reason, 
one can ignore the velocity-dependent terms in the xc force (eq.~(\ref{fxc})), 
and the difference between $\vec r$ and the ``retarded position" $\vec R$ of
the fluid element. Similarly, the apparent ambiguity  of whether the density
entering the viscosity coefficients in eq.~(\ref{stressnl}) should be evaluated
at time $t$ or at some earlier time $t'$, is  resolved by noting that the
difference $n(\vec r,t')-n(\vec r,t) = \int_{t'}^t \vec \nabla \cdot \vec
j(\vec r, \tau)d\tau $   generates a higher order gradient correction,
provided that the range of times which contribute significantly to the
integral in eq.~(\ref{stressnl}) is essentially finite.

The simple form of eq.~(\ref{axcnl})  is justified by our basic assumption
that 
the gradients of the density and velocity fields  be small. By using the full
expression (\ref{fxc}) for the electromagnetic force on the left hand side of
eq.~(\ref{axcnl}), and by replacing  $(\vec r,t') \to (\vec R(t'\vert
\vec r, t),t')$ on the right hand side of eq.~(\ref{stressnl}), the 
approximation can be systematically improved, so as to  satisfy the local
``zero 
force" theorem to {\it all}  orders in the gradients. 
A straightforward generalization allows one to construct an approximation
that also satisfies the local ``zero torque" theorem.

Finally,  we wish to comment on the condition  $k,q \ll 
\omega/v_F$ which defines the limit of validity of our approximate
treatment of $\vec a_{xc}$.  This restriction is forced on us by the
analytic structure of the functions $f^h_{xcL(T)}(k, \omega)$  which are
singular along the line $\omega = k v_F$, thus limiting the radius of
convergence of the small-$k$ expansion. The condition is reasonably
well satisfied at the characteristic frequencies of collective
excitations, but it  becomes increasingly restrictive as we lower the
frequency, entering the domain of single  electron-hole excitations.
Because the region  $\omega/kv_F \to0$ is not analytically
connected to the 
region $k \to 0$,  it seems unlikely that any local
density approximation can provide a physically sound description of
dynamical exchange and correlation in the electron-hole excitation 
region.     

In conclusion,  we have developed a consistent
and physically transparent formulation of nonlinear time-dependent
density functional theory beyond the ALDA.  The main physical
manifestation of dynamical exchange and correlation effects is the
appearance of viscosity and  dynamical shear coefficients in the
electron fluid.  This formulation allows  a first-principle
calculation of the {\it linewidth} of high-frequency collective
excitations, which are  not  Landau-damped, and would
therefore appear as sharp $\delta$-functions within the
ALDA. As a first application,  we
have obtained a compact expression for the linewidth $\Gamma$ of such
collective modes, to first order in the viscosity coefficients.  The
result is
 \begin {equation}
\Gamma =  \left| { {\rm Re} \sum_{i,k} \int d\vec r  u_i^\ast(\vec
r,\omega) 
\nabla _k \sigma_{xc,ik}(\vec r,\omega) \over m \int d\vec r  n_0(\vec r)
\vert \vec u(\vec r,\omega) \vert^2} \right|,
\label {linewidth} \end {equation} where $\vec u(\vec r)$ is the
velocity field of the collective mode calculated within the ALDA, and
$n_0(\vec r$) is the equilibrium density.  Detailed applications of this
formula will be reported elsewhere.

This work was supported by NSF grant No. DMR-9403908 and by a Research Board
Grant from the University of Missouri. S.C. acknowledges support from 
a travelling scholarship from Scuola Normale Superiore.  We acknowledge  useful
discussions with Walter Kohn, Hardy Gross, John Dobson, and Mario Tosi.


\begin{references}
\bibitem{Footnote1} We work in a gauge in which the dynamical external
scalar potential is zero.  In general,  a scalar potential $V(\vec r,t)$
is equivalent, by gauge invariance, to a longitudinal vector potential
$\vec a(\vec r,t) = c \int^t \vec \nabla V(\vec r, \tau) d \tau$.
\bibitem {Footnote0}  The limit $\omega \to 0$ here means $\omega\ll E_F$
while the condition 
$\omega/v_F\gg k,q$ is still satisfied. At finite temperature $T$, we must
also require 
$\omega \gg 1/\tau$  where  $\tau \sim T^{-2}$ is the collisional
relaxation time.  Similarly, if $\tau_{ext}$ is the relaxation time due to
collisions with external impurities, we require $\omega \gg 1/\tau_{ext}$.
\end{references}
\end {document}